\documentclass[%
 reprint,
 amsmath,amssymb,
 aps,
]{revtex4-1}

\usepackage{graphicx}% Include figure files
\usepackage{dcolumn}% Align table columns on decimal point
\usepackage{bm}% bold math
\usepackage{siunitx}
\DeclareSIUnit\wn{\raiseto{-1}\cm}
\DeclareSIUnit\rate{\cubic\centi\meter\per\second}
\begin{document}

\preprint{APS/123-QED}

\title{Rotational state-changing collisions between N$_2^+$ and Rb at low energies}

\author{A. D. D\"orfler}
 \affiliation{Department of Chemistry, University of Basel, Klingelbergstrasse 80, 4056 Basel, Switzerland}
 \author{E. Yurtsever}%
\affiliation{Department of Chemistry, Ko\c{c} University, Rumelifeneriyolu, Sariyer, TR-34450, Istanbul, Turkey}
\author{P. Villarreal}
 \affiliation{Instituto de F\'isica Fundamental, IFF\,-\,CSIC, Serrano 123, E-28006 Madrid, Spain}
 \author{T. Gonz\'alez-Lezana}
 \affiliation{Instituto de F\'isica Fundamental, IFF\,-\,CSIC, Serrano 123, E-28006 Madrid, Spain}
\author{F. A. Gianturco}\email{francesco.gianturco@uibk.ac.at}
\affiliation{Institut f\"ur Ionen Physik und Angewandte Physik, Leopold Franzens-Universit\"at, Technikerstrasse 25, A-6020, Innsbruck, Austria
}
\author{S. Willitsch}\email{stefan.willitsch@unibas.ch}
\affiliation{Department of Chemistry, University of Basel, Klingelbergstrasse 80, 4056 Basel, Switzerland
}
 
\bigskip

\date{\today}

\begin{abstract}
We present a theoretical study of rotationally elastic and inelastic collisions between molecular nitrogen ions and Rb atoms in the sub-Kelvin temperature regime prevalent in ion-atom hybrid trapping experiments. The cross sections for rotational excitation and de-excitation collisions were calculated using quantum-scattering methods on ab-initio potential energy surfaces for the energetically lowest singlet electronic channel of the system. We find that the rotationally inelastic collision rates are at least an order of magnitude smaller than the charge-exchange rates found in this system, rendering inelastic processes a minor channel under the conditions of typical hybrid trapping experiments.
\end{abstract}

\maketitle

\section{\label{sec:intro}Introduction}

Over the past ten years, the development of methods for the simultaneous confinement of ultracold atoms and ions has progressed rapidly enabling the exploration of interactions between charged and neutral particles in the sub-Kelvin temperature regime  \cite{haerter14a,sias14a,willitsch15a,willitsch17a,tomza19a}. Typical "hybrid" trapping experiments combine radiofrequency ion traps for the trapping and cooling of atomic and molecular ions \cite{willitsch12a} with magneto-optical \cite{grier09a,hall11a,rellergert11a,goodman12a,haze13a}, magnetic \cite{zipkes10a} or optical \cite{schmid10a,meir16a,joger17a} traps for the confinement of ultracold atoms, typically alkali species such as Rb, Ca or Li. Early studies focussed on elastic, inelastic and reactive collisions between atomic ions and neutral atoms and uncovered an astonishingly rich and diverse chemistry for these seemingly simple diatomic collision systems. Light-assisted processes on electronically excited collision channels were found to play a major role in many systems \cite{hall11a,sullivan12a} with non-adiabatic and radiative charge exchange as well as radiative association forming dominant reactive channels \cite{dasilva15a}. These reactive processes usually compete with elastic and inelastic collisions \cite{ratschbacher12a,sikorsky18a,sikorsky18b} which change the kinetic and internal energies of the scattering partners, but leave their chemical identity intact.

Most recently, a new focal area has developed in the field which aims at introducing molecular ions into hybrid experiments in order to study interactions between cold molecular ions and neutral atoms \cite{hall12a,rellergert13a,puri17a,doerfler19a}. The increased complexity of these collision systems opens up possibilities for more complex collision dynamics \cite{doerfler19a} and for additional reactive processes involving the formation and breaking of chemical bonds \cite{puri19a} which often compete with one another. The present article is the third in a series of a combined experimental and theoretical study aimed at exploring the role and competition of different collisional processes in the prototypical N$_2^+$ + Rb system in the cold regime. While our previous reports focussed on radiative association \cite{gianturco19a} and non-adiabatic charge exchange \cite{doerfler19a} in this system, the present study theoretically investigates the role of rotationally inelastic collisions using quantum scattering calculations on ab-initio potential energy surfaces (PES). 

As the main result of the present study, we find that the rate coefficients for rotationally inelastic collisions are at least an order of magnitude smaller than those for non-adiabatic charge exchange established in Ref. \cite{doerfler19a} at the collision energies relevant for hybrid trapping experiments. One can therefore conclude that inelastic processes only play a minor role under typical experimental conditions. The implications of the present findings for other cold molecular-ion alkali-atom collision systems are also discussed.

\section{\label{sec:comp}Computational Methods}

\subsection{\label{ssec:abini} Potential energy surface}

\begin{figure}[htpb!]
    \centering
    \includegraphics[width=1\columnwidth]{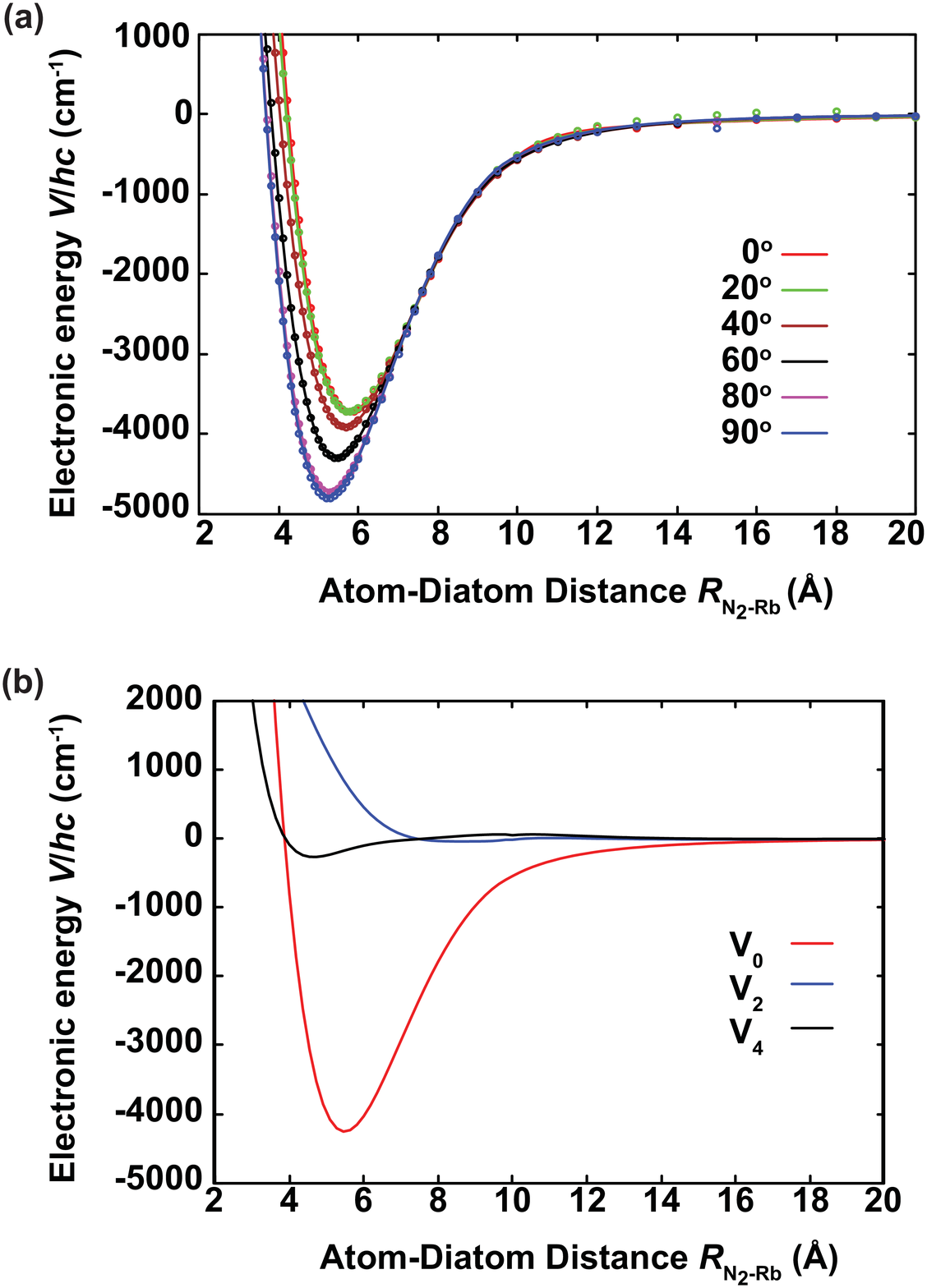}
    \caption{(a) One dimensional cuts of the potential-energy surface (PES) of the singlet channel of the N$^+_2~(^2\Sigma^+_g)$+Rb~$(5s~^2S_{1/2})$ collision system as a function of the N$_2^+$-Rb distance $R$ for different orientation angles $\theta$. The symbols represent the ab-initio points, lines correspond to fits of the ab-initio data to a Legendre expansion up to fourth order. (b) Radial expansion coefficients $V_\lambda, \lambda=0,2,4,$ of the Legendre expansion of the PES in (a) as a function of $R$. See text for details.}
    \label{fig:poten}
\end{figure}

The potential-energy surface (PES) used in the present scattering calculations is the same as in our previous study \cite{gianturco19a}, where details on the relevant computational methods can be found. Fig.~\ref{fig:poten}~(a) shows one-dimensional cuts through the PES of the singlet channel from Ref.~\citep{gianturco19a} as a function of the N$_2^+$-Rb distance $R$ at different N$_2^+$-Rb orientation angles $\theta$ with the N-N bond length fixed at its equilibrium value of $r=1.178$~\AA. 

To prepare the PES as input for the multichannel quantum scattering calculations outlined below, we fitted the ab-initio points in Figure~\ref{fig:poten}~(a) to a Legendre expansion up to fourth order \cite{gianturco19a}:
\begin{equation}
V(R,\theta) = \sum_{\lambda=0}^{\lambda_{max}}V_{\lambda}(R)P_{\lambda}(\cos\theta).
\label{eq:potlam}
\end{equation}
Here, $R$ is the distance between the centers-of-mass of the
collision partners and $\theta$ is their orientation angle, while $P_\lambda(\cos\theta)$ are Legendre polynomials. The expansion coefficients $V_\lambda (R)$ describe the relative strength of the anisotropy of each multipolar term. Because N$_2^+$ is a homonuclear diatomic molecular ion, only even terms in $\lambda$ occur.
%Here, $R$ is the distance between the centers-of-mass of the collision partners and $\theta$ is their orientation angle. The expansion coefficients $V_{\lambda}(R)$ describe the relative strength of the anisotropy of each multipolar term and are obtained from the computed potential $V(R,\theta$) according to
%\begin{equation}
% V_{\lambda} = \frac{1}{2\lambda+1}\int_{-1}^{1}V(R,\theta)P_{\lambda}(\cos\theta) \,d(\cos\theta),
%\label{eq:vlambda}\\[8pt]
%\end{equation}
%where $P_{\lambda}(\cos\theta)$ are the corresponding Legendre polynomials. Because N$_2^+$ is a homonuclear diatomic molecular ion, only even terms in $\lambda$ occur. 

Fig.~\ref{fig:poten}~(b) shows the dominant $V_{\lambda}$ coefficients with $\lambda=0\text{,}2\text{,}4$. Higher-order terms are not depicted because their contributions are shielded by the steep repulsive potential generated by the $\lambda=2$ term below $R=\SI{4}{\AA}$. The term with $\lambda=0$ corresponds to the isotropic polarizability $\alpha_0$ of the potential, the $\lambda=2$ term represents the  dipolar polarizability $\alpha_2$. Together they describe the polarizability as $\alpha_0P_0+\alpha_2P_2(\cos\theta)$. Both terms scale with $R^{-4}$ for $R\rightarrow\infty$. The interaction between the molecular ion and the interacting Rb atom is therefore described by the anisotropic forces at short distances up to the repulsive wall and by the attractive long-range forces caused by the molecular charge polarizing the impinging atom. The anisotropy of the interaction which is represented by the $\lambda=2,4$ terms is the driving force for rotationally inelastic collisions.

Within the interval 2.5 \AA \, $\le R \le$ 20 \AA, the potential curves of Fig. \ref{fig:poten} were fitted to the expansions
\begin{align}
\label{eq:three}
V(R,{\theta})=\begin{cases}
\sum\limits_{n=0}^6 p_n(\theta)\exp\lbrack-n\alpha(\theta)(R-R_0(\theta))\rbrack, & R\leq R_c \\
\sum\limits_{n=4}^8 \frac{C_{n}(\theta)}{R^{n}}, & R>R_c.
\end{cases}
\end{align}
where the $p_n(\theta)$ are expansion coefficients, $R_0(\theta)$ is a fit parameter approximating the position of the minimum of the potential at a given orientation angle $\theta$ and the $C_n(\theta)$ are the long-range multipolar coefficients. The matching point of the two expansions was chosen at $R_c=$10 \AA.
At this point, the long-range interaction already behaves properly at long distances where it should provide the leading term of the long-range interaction which, in this system, is of the form \ $-1/R^4$.   

From the averaged value $C_4=(C_4(0^\circ)+2C_4(90^\circ))/3$, we obtained the isotropic dipole polarizability of Rb, $\alpha^0_{Rb}=-2C_4\approx$ 46.7 \AA$^3$. This value is close to the experimental result for isolated Rb atoms ($\alpha_0=48.7$~\AA$^3$~\cite{hall10a}). 

\begin{table*}[!t]
  \begin{center}
  \begin{tabular}{l|rrr}
&$V(R,0^\circ)$\phantom{aaaaa} & $V(R,90^\circ)$\phantom{aaaaa} & $V(R,40^\circ)$\phantom{aaaa} \\  
\hline                                                                                                        
    \tabularnewline
\hline
\tabularnewline
$R_0$(\AA)        &5.8&5.25&5.72\\
$\alpha$(\AA$^{-1}$) &0.034468&0.034468&0.034468\\
$p_0$(cm$^{-1}$)  &-1756256.63&-262228016.79&-267045622.28\\
$p_1$(cm$^{-1}$)  &12544629.15&1657141129.59 &1703859985.34\\
$p_2$(cm$^{-1}$)  & -30069857.17&-4348781360.99&-4518541081.52\\
$p_3$(cm$^{-1}$)  &26102330.12&6062945894.88&6372877530.25\\
$p_4$(cm$^{-1}$)  &3701700.96&-4733245170.05&-5039612014.95\\
$p_5$(cm$^{-1}$)  & -17639551.38&1960411915.99&2117708797.62\\
$p_6$(cm$^{-1}$)  &7113286.78&-336249184.65&-369251509.65\\
$C_4$(cm$^{-1}$\AA$^4$)  &\phantom{aaa.}-2437572.32 & -2869054.88 & -2653313.60\\
$C_5$(cm$^{-1}$\AA$^5$) & -308701429.10&-535117.89&-32081850.31\\
$C_6$(cm$^{-1}$\AA$^6$)\phantom{}   & 5063017522.48& -113620695.71&242334325.22 \\
$C_7$(cm$^{-1}$\AA$^7$)\phantom{}   &-- &-- &-- \\
$C_8$(cm$^{-1}$\AA$^8$)\phantom{}   & -225693428343.86& -11575979814.12&-22768354089.67\\
\hline
  \end{tabular}
  \end{center}
  \caption{Parameters for the analytic description of the interaction potential $V(R,\theta)$ according to Eq. (\ref{eq:three}) at three orientation angles $\theta=0^\circ, 90^\circ$ and $40^{\circ}$. See text for details.} 
\label{tab:param}
\end{table*}

The radial coefficients up to fourth order were obtained from a collocation procedure considering the potential curves at $\theta=0, 40$ and $90^\circ$ to obtain the first three coefficients $V_0(R), V_2(R)$ and $V_4(R)$ of the expansion Eq. (\ref{eq:potlam}):
\begin{align}
  \label{eq:three_bis}
V_0(R)=&V(R,0^\circ) - V_2(R)- V_4(R),\nonumber\\
V_2(R)=&[0.694~ V(R,0^\circ)+0.625~ V(R,40^\circ) \nonumber \\
 &-1.319~ V(R,90^\circ)]/1.5911,\nonumber\\
V_4(R)=&[V(R,0^\circ)- V(R,90^\circ) - 1.5~ V_2(R)]/0.625.
\end{align}
The analytical fits are compared to the ab-initio data in Fig.\ref{fig:poten}~(a). As can be seen, the Legendre expansion up to fourth order represents the potential adequately. In Fig. \ref{fig:poten} (b), the expansion coefficients $V_\lambda(R)$ are plotted as a function of the inter-particle separation $R$. The numerical values of the fit parameters are listed in Tab.~\ref{tab:param}. The magnitude of the isotropic coefficient $V_0$ significantly exceeds the one of the anisotropic contributions $V_2$ and $V_4$ indicating that the potential is dominated by the strong polarizability of the multi-electron atomic partner, as can be expected from a non-polar molecular ion interacting with a neutral atom. From this finding, one can already expect elastic processes to dominate over inelastic ones in the present system. Moreover, the repulsive part of the $V_2(R)$ curve effectively shields large parts of the non-vanishing sections of the $V_4(R)$ potential so that it can be expected that the major inelastic processes will involve changes of two units of rotational angular momentum, i.e., $\Delta N=\pm2$. In this work, $N$ denotes the rotational quantum number of N$_2^+$, $j$ is the quantum number of the total molecular angular momentum without nuclear spin and $J$ is the quantum number of the total angular momentum of the collision system (excluding nuclear spin).

\subsection{\label{ssec:aspin} Multichannel quantum scattering calculations}

In order to obtain rotationally inelastic state-to-state integral scattering cross-sections (ICS), we have employed the ASPIN quantum-scattering code described in Ref. ~\cite{duran08a}. In the scattering calculations, the atom was treated as a structureless particle while the molecule was incorporated as a rigid rotor with singlet or doublet electron spin. 

The integration of the scattering equations for collision energies up to $E_{\mathrm{coll}}/k_{\mathrm{B}}=\SI{1100}{\kelvin}$ extended out to $R=220$~\AA \ using a total of 3000 steps. The maximum number of the total angular-momentum quantum number used in the calculations was $J_{max}=350$, while the maximum number of rotational channels employed was 23. For calculations at low collision energies up to $E_{\mathrm{coll}}/k_{\mathrm{B}}=\SI{70}{\milli\kelvin}$, the integration of the scattering equations extended out to $R=15000$~\AA \  using a total of 30000 steps and a maximum total angular-momentum quantum number of $J_{max}=18$, while the maximum number of rotational channels employed was 23. We expect that the final scattering observables are numerically accurate to about 5 percent of their reported values.

\section{\label{sec:anres} Results and discussion}

\begin{figure}[htpb!]
    \centering
    \includegraphics[width=1\columnwidth]{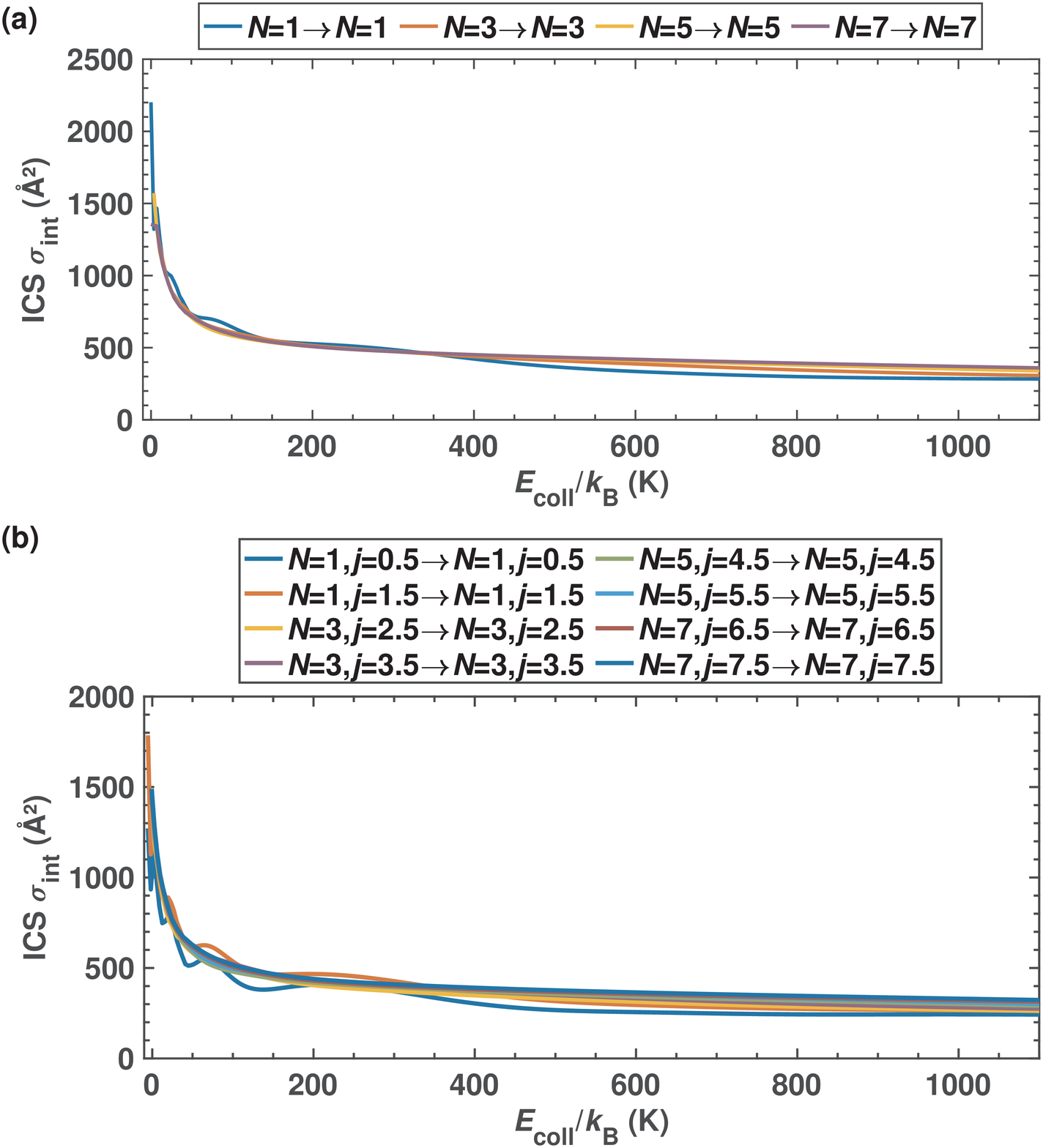}
    \caption{Computed rotationally elastic integral cross sections (ICS) for  N$_2^+$ + Rb by treating the molecular ion as a pseudo-singlet-spin species in (a) and as as a doublet-electron-spin species in (b). }
    \label{fig:el-ics}
\end{figure}

ICS for rotationally elastic collisions are reported in Fig. \ref{fig:el-ics}. In the results shown in Fig.~\ref{fig:el-ics} (a), the N$^+_2$ moiety has been treated as a species with pseudo-singlet electron spin, i.e., it was treated as a closed-shell system and spin-rotation coupling was neglected. Elastic cross sections are reported for the lowest rotational levels $N=1\text{,}3\text{,}5\text{,}7$ of the molecular ion up to collision energies of $E_\text{coll}/k_\text{B}=1000$~K. 
The computed ICS show a monotonic decrease with increasing collision energy which is a general behaviour often observed in elastic cross sections. 

The computational validity of treating N$^+_2$ as a pseudo-singlet rather than a doublet electronic state, the latter requiring the incorporation of spin-rotation coupling in the scattering calculations, is justified by comparing the spin-rotation resolved elastic cross sections in Fig.~\ref{fig:el-ics} (b) with the results for the pseudo-singlet in panel (a). The two sets of cross sections agree well with each other and also show a consistent energy dependence. The marginal differences which we see between the two different approximations are essentially negligible for the purpose of the present discussion of purely rotational effects in the scattering. The pseudo-singlet approach is producing physically reliable values for the quantities of interest here, i.e., rotationally resolved cross sections. We therefore treated the collisional dynamics within the pseudo-singlet approximation in order to reduce the computational costs.

\begin{figure}[htpb!]
    \centering
    \includegraphics[width=1\columnwidth]{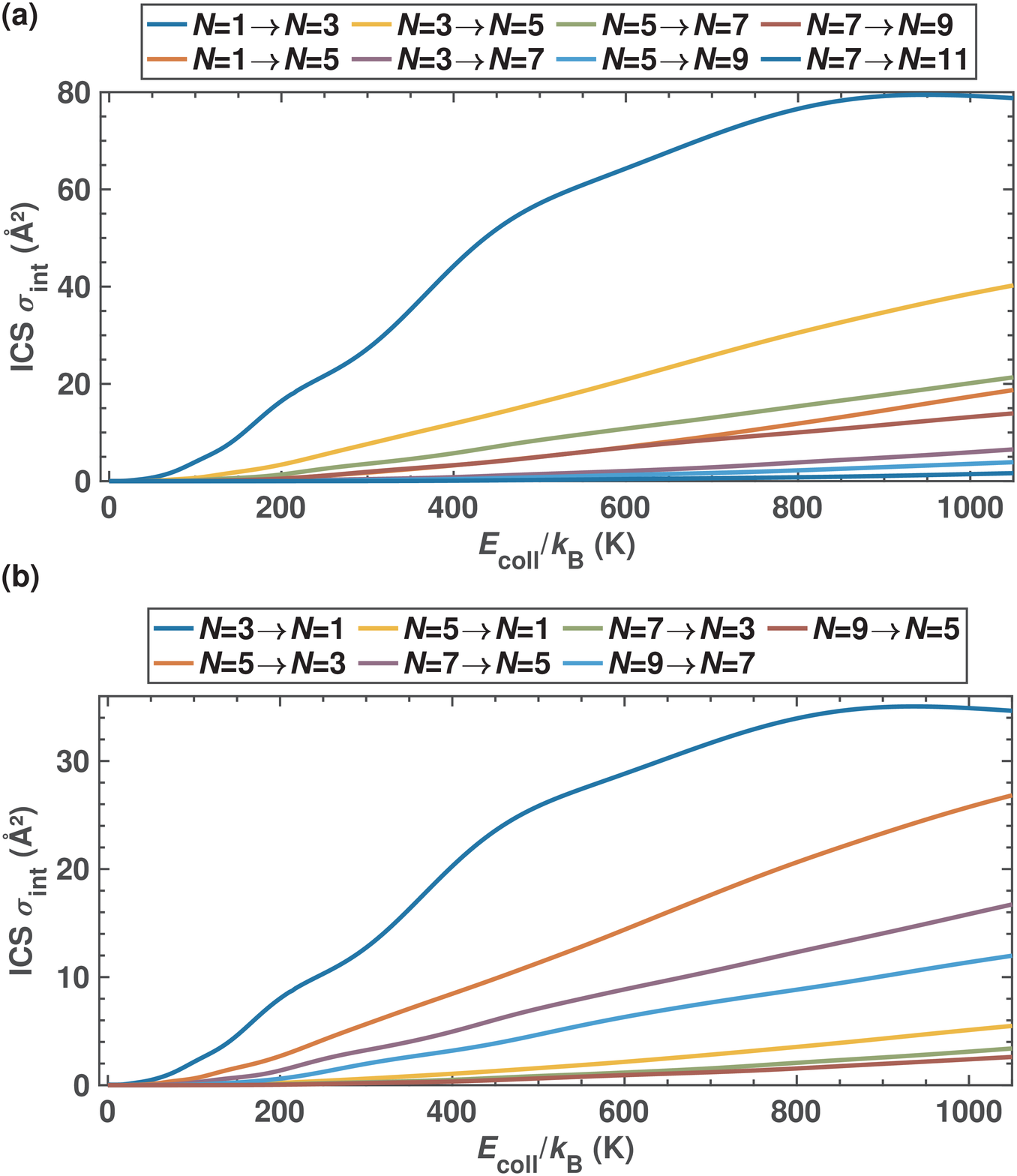}
    \caption{State-to-state excitation (a) and de-excitation (b) inelastic integral cross section (ICS) at collision energies up to 1100~K.}
    \label{fig:inel-ics}
\end{figure}

Figs. \ref{fig:inel-ics} (a) and (b) show ICS for rotational excitation and de-excitation processes, respectively, for the lowest rotational states in collisions between N$_2^+$ and Rb in the energy range up to $E_\text{coll}/k_\text{B}=1200$~K. Note that only state changes with $\Delta N$=even are allowed assuming the conservation of nuclear spin symmetry during the collision. We clearly see that over the entire range of collision energies examined, the rotational excitation and de-excitation cross sections show a fairly similar behaviour and monotonically increase with collision energy. Both processes exhibit rotationally inelastic cross section on the order of 1-100~\AA$^2$ \ in the interval $E_\text{coll}/k_\text{B}=100-1000$~K which is in line with the expected order of magnitude for such processes in an ionic system, see, e.g., Refs. \cite{gonzalez18a, hernandez18a} for comparison. The most distinctive difference lies in transitions involving the lowest rotational states, namely $N=1$ to 3 and $N=3$ to 5 and vice versa for which we find a factor of two difference between the ICS of the excitation and de-excitation processes. This is reasonable given the different energy gaps exhibited by the two processes. Note also that the cross sections involving changes in the rotational quantum number $\Delta N=\pm 4$ are about an order of magnitude smaller than those associated with $\Delta N=\pm 2$, reflecting the relative magnitude of the $V_{\lambda=2,4}(R)$ coefficients discussed above.

\begin{figure}[htpb!]
    \centering
    \includegraphics[width=1\columnwidth]{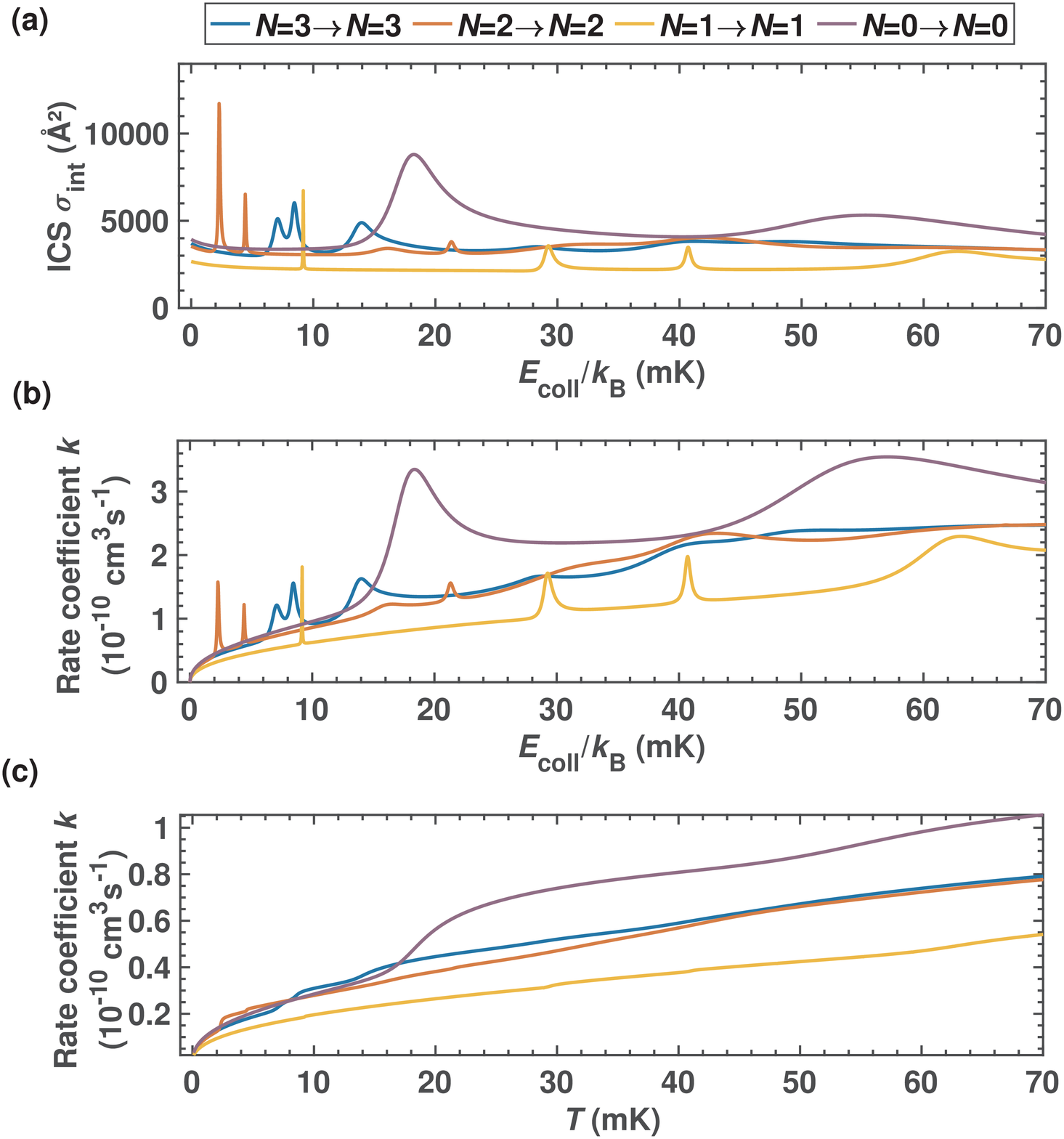}
    \caption{(a) Integral cross sections (ICS), (b) rate coefficients at an infinitely small spread of collision velocities and (c) thermal rate coefficients for N$_2^+$ + Rb elastic collisions in the cold regime.}
    \label{fig:el-ulow}
\end{figure}

We now turn to examining the cross sections in the cold regime at collision energies  around several tens of millikelvin which are particularly relevant for hybrid trapping experiments. Fig. \ref{fig:el-ulow} shows ICS (a) and rate constants $k$ (b,c) for rotationally elastic collisions. The rate constants in Fig. \ref{fig:el-ulow} (b)  were calculated assuming an infinitely narrow spread of collision energies in the experiment according to
\begin{equation}
    k=\sigma \sqrt{\frac{2E_{\mathrm{coll}}}{\mu}},\label{eq:rate}
\end{equation}
where $\sigma$ is the ICS and $\mu$ the reduced mass of the N$^+_2$\,-\,Rb collision. This situation would approximate an experiment with a high resolution in the collision energy \cite{eberle16a,puri18a}. By contrast, Fig. \ref{fig:el-ulow} (c) shows the thermal rate constants obtained by integrating the cross sections over a Maxwell-Boltzmann distribution of velocities at the relevant temperature. In this energy range, the cross sections exhibit marked variations attributed to scattering resonances. While these resonances also manifest themselves in the rate constants in experiments at high energy resolution (Fig. \ref{fig:el-ulow} (b)), the narrow features are almost completely smoothed out in the rate constants under thermal conditions (Fig. \ref{fig:el-ulow} (c)). 

The physical origin of the narrow resonances in the cross sections should be linked to the rather strong ionic interactions which are driving the present dynamics. The resonances observed in Fig. \ref{fig:el-ulow} could be shape resonances in which the system is intermittently trapped in a metastable state behind the centrifugal barrier. In addition, the strong coupling between scattering channels due to long-range ion-neutral interactions can also distort the rotational structure of the isolated molecular target creating virtual excitations to excited rotational states (Feshbach resonances) which can be populated during a short time interval during the collisional events. 
The narrow widths exhibited by the observed resonances, an indication of their lifetimes, are producing rather limited contributions to the thermal rate coefficients as can be seen in Fig. \ref{fig:el-ulow} (c). Hence, the computational evidence of the occurrence of several narrow resonant features near threshold energies would not translate into observable increases of the overall collision rates over the range of energies of interest studied in thermal experiments. Their observation would therefore require an experiment with a high collision-energy resolution as implied in the rate coefficients calculated in Fig. \ref{fig:el-ulow} (b) in which the resonance structures are preserved. 

Generally, the elastic rate coefficients are on the order of $10^{-10}$~cm$^3$s$^{-1}$ in this energy range, similar to other ionic systems, e.g., H$_2^+$ + He \cite{schiller17a}, but well below the Langevin rate coefficient $k_\text{L}=3.5\times10^{-9}$~cm$^3$s$^{-1}$ for N$_2^+$ + Rb \cite{doerfler19a}.

\begin{figure}[htpb!]
    \centering
    \includegraphics[width=1\columnwidth]{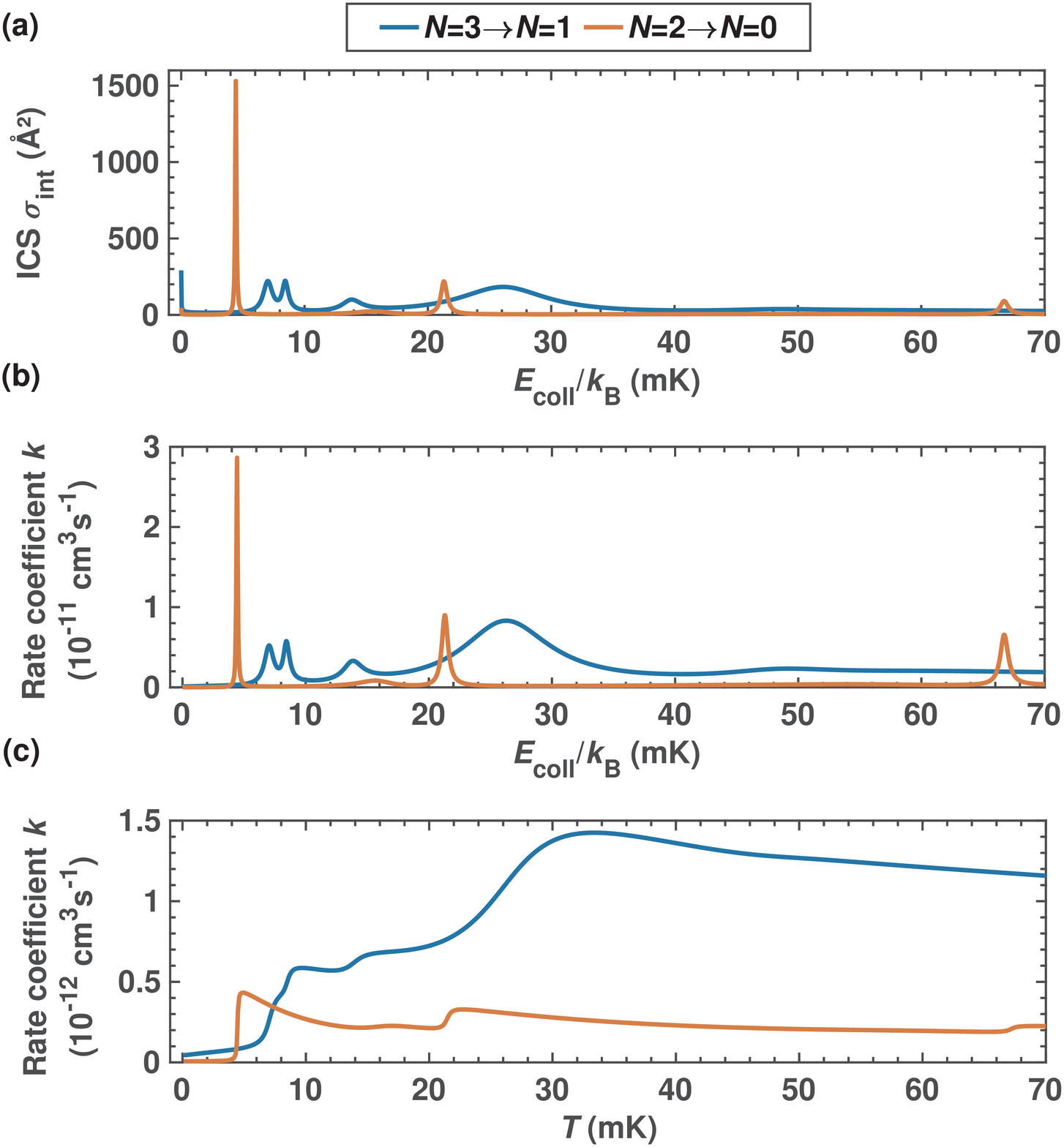}
    \caption{(a) Integral cross sections (ICS), (b) rate coefficients at an infinitely small spread of collision velocities and (c) thermal rate coefficients for N$_2^+$ + Rb rotationally inelastic collisions in the cold regime.}
    \label{fig:inel-ulow}
\end{figure}

Fig. \ref{fig:inel-ulow} shows rotational de-excitation cross sections (a) and rate coefficients (b,c) across the same range of collision energies. Note that rotational excitations are energetically forbidden at these low collision energies. The cross sections again show pronounced modulations as a function of the collision energy pointing to scattering resonances. Some of the strongest resonances, e.g., the one observed in the $N=2\rightarrow N=0$ channel around $E_\text{coll}/k_\text{B}=5$~mK, even manifest themselves in the thermal rate constants after thermal averaging in Fig.  \ref{fig:inel-ulow} (c). Except at the positions of the strongest resonances observed at high energy resolution, the rotational de-excitation rate coefficients are small with $k<10^{-11}$~cm$^3$s$^{-1}$ in the energy range shown. In particular, they are well below the Langevin limit for this system.

Together with the findings of our previous studies in Refs. \cite{gianturco19a, doerfler19a}, the present results on inelastic and elastic collision cross sections provide a comprehensive picture of the importance of the different types of collisional processes which can occur in the cold scattering of N$_2^+$ + Rb in their relevant ground states. The fastest process is established to be non-adiabatic charge transfer in the triplet scattering channel of N$_2^+~(X~^2\Sigma^+_g)$ with Rb~$(5s~^2S_{1/2})$ the rate constants for which were found to be on the order of $1.5\times10^{-9}$~cm$^3$s$^{-1}$ at collision energies in the range of tens of mK \cite{doerfler19a}. This value is only about a factor of 3 smaller than the Langevin limit. Charge exchange thus clearly dominates over elastic and inelastic collisions the rate coefficients of which were calculated here to be at least one to two orders of magnitude smaller. By contrast, the rates for radiative association to form the N$_2$Rb$^+$ complex were found to be at least five orders of magnitude smaller than the Langevin limit, rendering this process insignificant in experiments. This is in line with the experimental results of Ref. \cite{doerfler19a} in which charge transfer was indeed found to be the dominant chemical process. 

We argue that the dynamical picture which has been found here in N$_2^+$ + Rb can qualitatively be transferred to a wider range of cold collision systems consisting of small molecular ions and alkali atoms with a large difference in the ionization potentials of the two species. In these cases, the possible entrance channels of the collisions will likely be highly excited states of the collision complex in which opportunities may arise for non-adiabatic transitions around curve crossings with other excited states. In this scenario, charge exchange will likely be the kinetically dominating process, as in N$_2^+$ + Rb and most probably also O$_2^+$ + Rb \cite{doerfler19a}. Rotationally inelastic collisions, as studied here, and competing reactive processes such as radiative association \cite{gianturco19a} are likely to exhibit considerably smaller rates. Of course, this does not preclude the possibility of other fast chemical processes which may occur depending on the specific system \cite{puri17a}. 

\section{Conclusions}
\label{sec:conc}

In this work we have calculated the rate coefficients of elastic and rotationally inelastic collisions in the N$_2^+~(X~^2\Sigma^+_g)$\,+\,Rb~($5s$~$^2S_{1/2}$) system. Collision cross sections were obtained from multichannel quantum-scattering calculations performed on ab-initio potential-energy surfaces of the lowest singlet collision channel. The magnitude and the general trends observed for both the elastic and inelastic cross sections were found to be in line with similar ionic collision systems explored in previous studies. Notably, the elastic and inelastic collision rates were found to be considerably smaller than the charge-transfer reaction rates \cite{doerfler19a}. Thus, it can be expected that any inelastic processes will be dominated by reactive processes in this system and inelastic collisions will play a small role under typical experimental conditions. Similar scenarios are likely to be found over a wider range of collision systems involving small molecular ions and alkali atoms with a large difference in the ionization energies between the two species.

\bigskip

\section*{Acknowledgements}

The present work was supported by the Swiss National Science Foundation, grant nr. 200020\_175533, and the University of Basel. PV and TGL acknowledge the support of the Spanish MINECO through grant nr. FIS2017-83157-P.

%\bibliographystyle{apsrev4-1}
%\bibliography{Main-Aug19,newref}

%

\end{document}